\documentclass[fleqn,usenatbib]{mnras}

\usepackage{amssymb}
\usepackage{newtxtext,newtxmath}
\usepackage{siunitx}
\usepackage[utf8]{inputenc}
\usepackage[T1]{fontenc}
\usepackage{CJKutf8}
\usepackage{booktabs}
\DeclareRobustCommand{\VAN}[3]{#2}
\let\VANthebibliography\thebibliography
\def\thebibliography{\DeclareRobustCommand{\VAN}[3]{##3}\VANthebibliography}

\usepackage{tikz,xcolor,hyperref,colortbl}
\usepackage{graphicx}	
\usepackage{amsmath}	
\title[From Galaxy Zoo DECaLS to BASS/MzLS]{From Galaxy Zoo DECaLS to BASS/MzLS: detailed galaxy morphology classification with unsupervised domain adaption}
\usepackage{orcidlink}
\defcitealias{walmsley_galaxy_2021}{W+21}
\newcommand{\WM}{\citetalias{walmsley_galaxy_2021}}
\author[Ye et al.]{
Renhao Ye~\begin{CJK}{UTF8}{gbsn}(叶人豪)\end{CJK}~\orcidlink{0000-0002-2339-5581}$^{1,2}$,
Shiyin Shen~\begin{CJK}{UTF8}{gbsn}(沈世银)\end{CJK}~\orcidlink{0000-0002-3073-5871}$^{1,4}$,\thanks{Email: \href{mailto:ssy@shao.ac.cn}{ssy@shao.ac.cn}}
Rafael S. de Souza~\orcidlink{0000-0001-7207-4584}$^{3}$,\thanks{Email: \href{mailto:rd23aag@herts.ac.uk}{rd23aag@herts.ac.uk}}
\newauthor~Quanfeng Xu~\begin{CJK}{UTF8}{gbsn}(徐权峰)\end{CJK}~\orcidlink{0000-0002-9572-6212}$^{1,2}$,
Mi Chen~\begin{CJK}{UTF8}{gbsn}(陈宓)\end{CJK}~\orcidlink{0009-0006-4205-6450}$^{1,2}$,
Zhu Chen (\begin{CJK}{UTF8}{gbsn}陈竹\end{CJK}) $^{4}$,
Emille E. O. Ishida~\orcidlink{0000-0002-0406-076X}$^{5}$,
\newauthor~Alberto Krone-Martins~\orcidlink{0000-0002-2308-6623}$^{6,7}$ and 
Rupesh Durgesh$^{8}$
\\
$^{1}$Shanghai Astronomical Observatory, Chinese Academy of Sciences, 80 Nandan Road, Shanghai 200030, China\\
$^{2}$School of Astronomy and Space Science, University of Chinese Academy of Sciences, No. 19A Yuquan Road, Beijing 100049, China\\
$^{3}$Centre for Astrophysics Research, University of Hertfordshire, College Lane, Hatfield, AL10~9AB, UK\\
$^{4}$Shanghai Key Lab for Astrophysics, Shanghai Normal University, 100 Guilin Road, Shanghai 200234, China\\
$^{5}$LPCA, Université Clermont Auvergne, CNRS/IN2P3, F-63000 Clermont-Ferrand, France\\
$^{6}$Donald Bren School of Information and Computer Sciences, University of California, Irvine, CA 92697, USA\\
$^{7}$CENTRA/SIM, Faculdade de Ciências, Universidade de Lisboa, Ed. C8, Campo Grande, P-1749-016 Lisboa, Portugal\\
$^{8}$Independent Researcher\\
}
\date{Accepted XXX. Received YYY; in original form ZZZ}
\pubyear{2024}
\begin{document}
\label{firstpage}
\pagerange{\pageref{firstpage}--\pageref{lastpage}}
\maketitle
\begin{abstract}
The DESI Legacy Imaging Surveys (DESI-LIS) comprise three distinct surveys: the Dark Energy Camera Legacy Survey (DECaLS), the Beijing-Arizona Sky Survey (BASS), and the Mayall $z$-band Legacy Survey (MzLS).
The citizen science project Galaxy Zoo DECaLS 5 (GZD-5) has provided extensive and detailed morphology labels for a sample of \num{253287} galaxies within the DECaLS survey. This dataset has been foundational for numerous deep learning-based galaxy morphology classification studies. However, due to differences in signal-to-noise ratios and resolutions between the DECaLS images and those from BASS and MzLS (collectively referred to as BMz), a neural network trained on DECaLS images cannot be directly applied to BMz images due to distributional mismatch.
In this study, we explore an unsupervised domain adaptation (UDA) method that fine-tunes a source domain model trained on DECaLS images with GZD-5 labels to BMz images, aiming to reduce bias in galaxy morphology classification within the BMz survey. Our source domain model, used as a starting point for UDA, achieves performance on the DECaLS galaxies' validation set comparable to the results of related works. For BMz galaxies, the fine-tuned target domain model significantly improves performance compared to the direct application of the source domain model, reaching a level comparable to that of the source domain. We also release a catalogue of detailed morphology classifications for \num{248088} galaxies within the BMz survey, accompanied by usage recommendations.
\end{abstract}

\section{Introduction}

Galaxy morphology is a cost-effective proxy for assessing galaxy diversity and its physical properties. Originally proposed by \cite{hubble_extragalactic_1926}, the Hubble classification scheme organized local galaxies into elliptical, lenticular, spiral, and irregular categories. The morphology of a galaxy is not only an external expression of its structure but is also closely related to its stellar population \citep{gonzalez_delgado_califa_2015} and environment \citep{margoniner_photometric_2000, goto_morphological_2003}. With advances in imaging depth and resolution, more detailed morphology features such as spiral arms, dust lanes, bars, and tidal tails can be observed in these extragalactic galaxies, enabling more systematic studies of their physical properties. Previous literature has explored the correlation between the global physical properties of spiral galaxies and the number of their spiral arms \citep{hart_galaxy_2016,hart_galaxy_2017,porter-temple_galaxy_2022}, the strength of the bar and the quenching process \citep{kruk_galaxy_2017,geron_galaxy_2021}, the global morphology and bulge fraction \citep{kumar_growth_2022,kumar_study_2023}, and the merging stage and quenching pathway of galaxy mergers \citep{darg_galaxy_2010,weigel_galaxy_2017}.

The morphology of a galaxy is typically obtained through visual inspection. The Galaxy Zoo (GZ) project \citep{lintott_galaxy_2008} is a pioneering citizen science effort in which volunteers visually classify galaxy images into different morphology labels. GZ projects, including GZ1, GZ2, and GZ DECaLS \citep{lintott_galaxy_2011, willett_galaxy_2013, walmsley_galaxy_2021}, organized many morphology studies. By continuously collecting new contributions from volunteers, these GZ-based galaxy morphology catalogues have greatly facilitated astronomical galaxy morphology studies \citep{zhang_spin_2015,geron_galaxy_2021}. As more and more galaxies will be observed by the next generation of telescopes, classifying billions of galaxies through volunteers alone will be impossible. Aligned with the development of deep learning techniques in computer vision, GZ catalogues provide excellent training samples for supervised deep learning \citep{dieleman_rotation-invariant_2015, dominguez_sanchez_improving_2018, walmsley_galaxy_2020, seo_similar_2023}. Based on a series of GZ-related works, a model called \texttt{Zoobot} for detailed galaxy morphology classification was developed by \cite{walmsley_galaxy_2021, walmsley_towards_2022, walmsley_practical_2022, walmsley_rare_2023}, aiming to build a foundational model that can be applied to galaxies in other surveys.

The DESI Legacy Imaging Surveys (DESI-LIS) comprise three distinct surveys: the Dark Energy Camera Legacy Survey (DECaLS), the Beijing-Arizona Sky Survey (BASS), and the Mayall $z$-band Legacy Survey (MzLS), together producing a new generation of galaxy imaging dataset with superior depth and coverage compared to the Sloan Digital Sky Survey (SDSS). Among the three DESI-LIS surveys, the pixel scale of BASS is larger compared to DECaLS and MzLS, and different $g$-band filter efficiencies compared to DECaLS \citep{he_quasar_2022}, resulting in subtle and systematic differences in the image files from these surveys, which can be referred to as data shift. Predicting galaxy morphology labels from a survey with characteristics different from those used for training will likely lead to biased predictions if not properly considered \citep{huang_adversarial_2011,goodfellow_explaining_2015,pooch_can_2020}. \cite{xu_images_2023} have shown that there are systematic discrepancies in the latent space of common galaxies between the DECaLS and BMz surveys. Similarly, \cite{he_quasar_2022} find that the data shift from DECaLS to BMz affects the completeness of the predictions for quasi-stellar objects (QSOs). In addition, label shift may also influence visual inspection (True labels) of galaxy morphology (see more discussions in Sect.~\ref{sec: cp23}.)

To mitigate the impact of data shift, a straightforward approach is to train each dataset simultaneously \citep[e.g.][]{walmsley_galaxy_2023}. Alternatively, we can fine-tune the model employing transfer learning \citep{dominguezsanchez_transfer_2019, hannon_star_2023, tang_transfer_2019, ackermann_using_2018} or domain adaptation techniques \citep{xu_images_2023, ciprijanovic_deepadversaries_2022, ciprijanovic_deepastrouda_2023}. Transfer learning involves fine-tuning a pre-trained model on a specific downstream task, leveraging the knowledge acquired in the initial training phase. Domain adaptation (DA), a subset of transfer learning, addresses data shift by aligning embedding distributions or finding embeddings that are domain-invariant, ensuring the model generalises well across different domains. As a subset of DA, unsupervised domain adaptation (UDA) focusses on aligning invariant embeddings in datasets of different domains without collecting labels \citep{li_model_2020, huang_category_2022, wang_cross-domain_2023, li_unsupervised_2024}. 

Empirically, both supervised transfer learning and domain adaptation require less data compared to training a model from scratch \citep{tahmasebzadeh_schwarzschild_2023,euclid_collaboration_euclid_2024}, but obtaining sufficient labels for under-represented classes (e.g. minor merger, spiral galaxies with three arms) remains a challenge. We use UDA to predict the galaxy morphology classifications from DECaLS images to BMz galaxies. This approach is justified because a finite number of galaxies with GZD-5 labels exist in both the DECaLS and the BMz surveys and the sample size in both surveys is large enough to identify under-represented classes and align invariant embeddings in the latent space. The UDA technique is well suited for datasets from the same physical domain, such as different galaxy surveys with different instruments. 

Our training strategy through UDA involves a two-step process. First, we train a source domain model using DECaLS images and GZD-5 labels. After this source domain training, we fine-tune the source domain model using \num{248088} unlabelled galaxies in the BMz survey, which are referred to the target domain. Importantly, this fine-tuning process leverages only unlabelled data from the target domain, without utilising any labelled galaxies that might be common to both the source and target datasets. Finally, we evaluate the target domain model's performance on \num{3618} labelled BMz galaxies in the overlapping region between DECaLS and BMz. This work represents a testbed study for the implementation of established deep learning models in new galaxy survey samples such as the Chinese Space Station Telescope \citep[CSST, ][]{gong_cosmology_2019}, Euclid \citep{euclid_collaboration_euclid_2022}, and the Vera C. Rubin Observatory Legacy Survey of Space and Time \citep[LSST, ][]{ivezic_lsst_2019}. 

This paper is structured as follows. In Sect.~\ref{sec: data}, we introduce the galaxy sample in the DESI-LIS and GZD-5 labels. In Sect.~\ref{sec: model}, we present the architecture of the models of the source and target domains and details of the experiment. In Sect.~\ref{sec: results}, we evaluate the performance of the model in both the source and the target domains. In Sect. \ref{sec: use}, we introduce the usage of the catalogue that we released. Finally, conclusions are drawn in Sect.~\ref{sec: conclusion}.

\section{Dataset}

\label{sec: data}
\subsection{Main galaxy sample in DESI-LIS}
\label{subsec: mgs}
\begin{table}
    \caption{Survey parameters of DESI-LIS}
    \centering
    \resizebox{\columnwidth}{!}{
    \begin{tabular}{ccccc}
    \hline
        Survey & Instrument & Bands & Area & Pixel scale \\
        \hline
        DECaLS & Blanco 4m/DECam & $grz$ & $\sim9000^\circ$& 0\farcs262 \\
        BASS & Bok 2.3m/90Prime & $gr$ & $\sim5000^\circ$& 0\farcs454 \\
        MzLS & Mayall 4m/MOSAIC-3 & $z$ & $\sim5000^\circ$& 0\farcs262 \\
    \hline
    \end{tabular}}
    \label{tab: survey}
\end{table}

The Main Galaxy Sample (MGS) for spectroscopy in the Sloan Digital Sky Survey (SDSS) consists of galaxies with $r$-band Petrosian magnitudes $m_r\leq 17.77$ \citep{strauss_spectroscopic_2002}. MGS has served as a milestone sample for studying the physical properties of low-redshift galaxies \citep{blanton_luminosity_2001,shen_size_2003,kauffmann_stellar_2003, baldry_color_2004,yang_galaxy_2007}. Additionally, the GZ1 project has provided a morphology classification baseline (e.g. spiral, elliptical, merger) for the MGS. The GZ2 project extends this with more detailed classifications, including features such as the number and tightness of spiral arms and whether the galaxy is edge-on.

As a new-generation sky survey, the DESI-LIS \citep{dey_overview_2019} provides imaging results deeper than those of the SDSS across the sky in roughly \num{20000} deg$^2$ in the $grz$ bands. For MGS in the DECaLS survey, the GZD-5 project has provided detailed volunteer morphology votes for \num{253287} of them (\citealt{walmsley_galaxy_2021}, hereafter \WM).\footnote{We don't use the prediction results of \num{314000} galaxies from \WM.}

For the MGS in the BMz survey, the dataset for UDA, we follow the GZD-5 selection criterion, i.e. $z\sim 0.15, m_r<17.77$, and Petrosian radius $> 3$ arcsec, resulting in $\num{248088}$ galaxies, which are referred to the newly selected galaxies as BMz galaxies. This selection criterion ensures that the physical domain of BMz galaxies are the same as those of the DECaLS galaxies. Additionally, there are \num{3618} common DECaLS/BMz galaxies in the overlap footprints between the DECaLS and BMz surveys (around $32^\circ < \delta < 34^\circ$).

\begin{figure}
\includegraphics[width=\columnwidth]{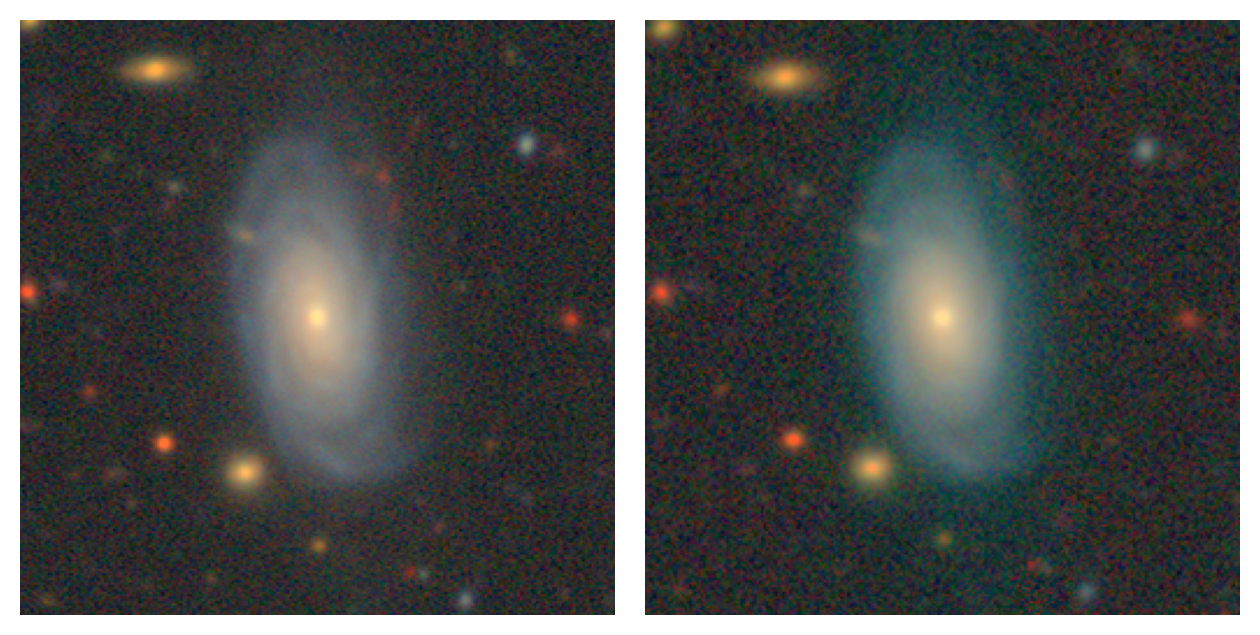}
    \caption{Composite images (consisting of $grz$ bands) of a randomly selected spiral galaxy in DECaLS (left) and BMz (right). Both images are being processed using the same arcsinh stretching method as the DESI Legacy Survey Viewer.}
    \label{fig: shift}
\end{figure}
\subsection{Stamp images of DECaLS/BMz galaxies}
As shown in Table~\ref{tab: survey}, the pixel scale of BASS (0\farcs454 arcsec/pixel) is larger compared to that of DECaLS and MzLS (0\farcs262 arcsec/pixel). However, in the DESI-LIS data pipeline, images from BASS were resampled to match the pixel scale of DECaLS and MzLS. This operation results in a different background noise distribution of BMz images compared to DECaLS images, as demonstrated in Fig. \ref{fig: shift}. We obtain the FITS image stamps (256 $\times$ 256 pixels with 0\farcs262 arcsec/pixel) in the $grz$ bands for both DECaLS and BMz galaxies using the cut-out service from DESI-LIS Data Release 9.\footnote{\url{www.legacysurvey.org/dr9/description/}} For common galaxies in the overlapping footprints, we acquired images from both surveys.\footnote{There is a small fraction of galaxies that will be outside the boundary of the cutout stamps, which are mainly nearby local galaxies with z<0.01.}

\subsection{Labels}
\label{subsec: dataset}
We train the source domain model on the GZD-5 volunteers' votes that has been modified for volunteers' weighting and redshift debiasing for \num{253287} galaxies.\footnote{As shown by \WM, these volunteer votes have been corrected for redshift bias and volunteer's bias.} Following \WM, we excluded the question `Do you see any of these rare features?' in the GZD-5 decision tree. As a result, there are 10 questions with a total of 34 features. To avoid ambiguity, the `answer' chosen by the volunteer is equivalent to this morphology `feature' of the galaxy. The final decision tree is the same as in fig. 5 of \WM. For each galaxy, volunteers' votes for a given morphology question range from a few to several dozen. To ensure that the labels are informative for training the source domain model, we train only on questions with at least 3 volunteers' votes. Finally, we have about \num{249581} galaxies with more than 3 votes on a total of 10 morphology questions.
\section{Method}
\label{sec: model}
\begin{figure*}
    \centering
    \includegraphics[width=\textwidth]{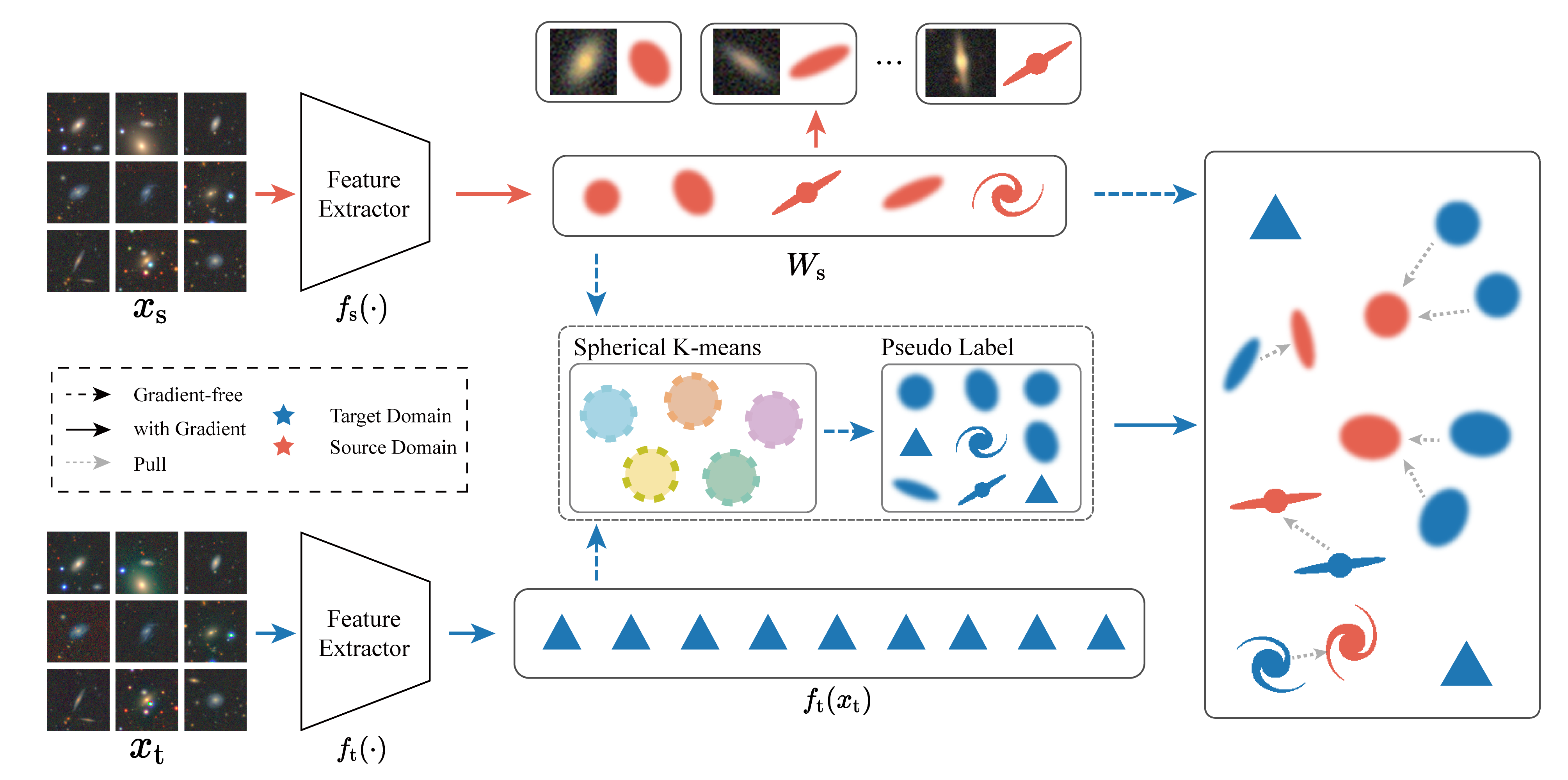}
    \caption{The schematic diagram of target domain training including cutout input $x_\text{s}$ from source domain, $x_\text{t}$ from target domain, the feature extractor $f_\text{s}(\cdot)$ and $f_\text{t}(\cdot)$, and the classifier $W_\text{s}$. Spherical K-means are used to obtain pseudo-labels. The triangle represents the feature embedding not assigned with a pseudo-label, and the galaxy-like shape represents the feature embedding of assigned morphology. Red colour represents the source domain feature embedding and blue represents the target domain.}
    \label{fig: diagram}
\end{figure*}
In this section, we detail our two-step training approach: (1) source domain model training on DECaLS images and GZD-5 votes, and (2) target domain model training on BMz galaxies. Although the first step mainly follows the methodology described in \WM, we have made some modifications, including using the raw FITS images without stretching and using a new neural network architecture. In the second step, we modify the UDA method in \cite{wang_cross-domain_2023} to fit our decision tree-based scenario, adapting the source domain model to the target domain. The diagram of the two steps is illustrated in Fig.~\ref{fig: diagram}.
\subsection{Source domain model}
\label{subsec: network}
Intuitively, we can train a model to predict the behaviour of GZ volunteers by treating each morphology question as a Multinomial distribution case. With a large number of volunteers, the vote fraction for each galaxy morphology feature approximates its true probability. However, because collecting enough volunteer votes for each galaxy is time consuming, we often have only a limited number of votes. To address this, we can model the responses of volunteers as sampling from a Dirichlet distribution and make the model predict the Dirichlet distribution of each morphology question, as introduced in \WM. Dirichlet distributions are parameterized by a positive value group $\boldsymbol{\alpha}=\{\alpha_1,\dots,\alpha_i,\dots,\alpha_n\}$, where $n$ is the dimension of the distribution (for example, how many answers a question has). 

Consider a question $q$ that has $m$ different features and a total of $N_q$ votes. We can easily calculate the vector of the fraction of votes $\boldsymbol{k_q}$ of each answer in question $q$. 
The source domain model aims to learn to predict the true probability $\boldsymbol{\rho_q}$, which is related to $\boldsymbol{k_q}$ and $N_q$ through the multinomial distribution $\operatorname{Multi}(\boldsymbol{k_q} \mid \boldsymbol{\rho_q}, N_q)$. Following \WM, we use the conjugate prior distribution of the Multinomial distribution, the Dirichlet distribution, $\text{Dir}(\boldsymbol{\alpha}_q)$, to predict $\boldsymbol{\rho}_q$.  The source domain model is then optimised by
\begin{equation}
    \mathcal{L}_\text{s}=-\log\sum_{q=1}^{10}\int \operatorname{Multi}(\boldsymbol{k}_q \mid \boldsymbol{\rho}_q, N_q) \operatorname{Dir}(\boldsymbol{\rho}_q \mid \boldsymbol{\alpha}_q) d \boldsymbol{\rho}_q.
\end{equation}
We then minimize $\mathcal{L}_\text{s}$, assuming a set of concomitant \footnote{During training, we optimize by summing the loss for all the question q, which assuming that all the question are depenent} $q = 10$ questions for efficiency. A detailed explanation of the Dirichlet distribution and the principle of this loss function can be found in \WM and \cite{walmsley_galaxy_2023}.

For the training set, we remove the galaxies in the overlapping footprint between DECaLS and BMz and divide the remaining galaxies into a train-valid set by a split of 80:20. The architecture of the model is EfficientNet-v2-s \citep{tan_efficientnetv2_2021} and we use the implementation in \textsc{torchvision}. For each image, we apply the following data augmentation techniques during training: (1) we randomly flip the image vertically or horizontally, each with a 50$\%$ probability, and (2) we randomly rotate the images anywhere from $0\sim 180^\circ$. We scale the model output scores following \WM with a sigmoid layer and multiply it by 100 and add 1 to obtain a range from 1 to 101, which meets the requirements of $\boldsymbol{\alpha}_q$. 

Our training process use an NVIDIA A100 80G GPU, using the AdamW optimiser \citep{loshchilov_decoupled_2019} with a learning rate strategy that starts at a maximum of 1e-2 and a minimum of 1e-6 and adapts according to a OneCycleLR scheduler \citep{smith_super-convergence_2018}. We set the batch size to 256 and the dropout rate to 0.3 to avoid overfitting. The beta parameters of optimiser are set to (0.9,0.99). We stop the training process after observing no further decrease in validation loss for 10 consecutive epochs. 

\subsection{Target domain model} 
\label{subsec: transfer}
Before describing the target domain model, we revisit the framework of our source domain model. Our source domain model comprises a latent embedding extractor $f_\text{s}(\cdot)$ and a classifier $W_\text{s}$. We denote all the 34 morphology features with indices $(q, m_q)$, representing the morphology feature (i.e. answer) $m_q$ of the question $q$. With this notation, $W_\text{s}$ is a set of weights for all latent embeddings, where $w_\text{s}^{q,m_q}$ represents the weight corresponding to the feature $m_q$. The scores, calculated by the product of the latent embeddings $f_\text{s}(x_\text{s})$ and the weights of the classifier $W_\text{s}$, represent the classification results before applying the Softmax function. If the model correctly classifies a morphology feature $m_q$, then the corresponding score should be the highest among all $m_q$ and result in low loss.

When it comes to the target domain (BMz) galaxies $x_\text{t}$, the latent embeddings $f_\text{s}(x_\text{t})$ from the source domain model have data shift and will show systematic bias. To address the data shift, we adopt UDA method from  \cite{wang_cross-domain_2023} to fine-tune $f_\text{s}(\cdot)$ to $f_\text{t}(\cdot)$ so that latent embeddings$f_\text{t}(x_\text{t})$ can be classified by the same $W_\text{s}$. Since the ground truth morphology feature $(q,\bar{m}_{q,i})$ is not available for individual galaxy, we generate (pseudo) labels for their latent embedding $f_\text{t}(x_{\text{t},i})$ by spherical K-means.

Before performing spherical K-means at the beginning of each epoch, we first fit the spherical K-means in the entire BMz training set by minimizing the cosine distance between $f_\text{t}(x_{\text{t},i})$ and the clustering centre $O^{q,\bar{m}_{q}}$ of the previous iteration, where $O$ are initialized by the weights of the classifier $W_\text{s}$. After fitting K-means, we generate pseudo-labels $\tilde{m}_{q,i}$ for galaxies that satisfies $f_\text{t}(x_{\text{t},i})\cdot O^{q,\bar{m}_{q}}$ > THRESHOLD , and use them for fine-tuning $f_\text{t}(\cdot)$. We set a conservative threshold (0.9-epoch * 0.02) because, in the initial stages of the unsupervised domain adaptation (UDA) process, the search region near $W_s$  must be carefully constrained. This is especially important in high-dimensional latent spaces, where an overly large region can cause instability or suboptimal results.

Specifically, the weights of the classifier $W_\text{s}$ if fixed and the $f_\text{t}(\cdot)$ is fine-tuned by the UDA loss function
\begin{equation}
\label{eq:uda_loss}
    \mathcal{L}_{\text{DA},i}=-\sum_{q=1}^{10}\log \text{Softmax} \left(f_\text{t}(x_{\text{t},i})^{\top} w_\text{s}^{q,\tilde{m}_{q,i}} / \tau\right),
\end{equation}
where $\tau$ is a temperature hyper-parameter that controls the sharpness of the Softmax output, making the probabilities either flatter or more concentrated \citep{he_momentum_2020}.

In the UDA model, we set the temperature $\tau=0.05$ and a learning rate of 1e-6 using the AdamW optimiser with beta = (0.9,0.99) and a batch size of 640. We stop training after no improvement for 5 consecutive epochs. 

\begin{table}
    \caption{Classical performance metrics for the GZD-5 validation set. Each row represents a question. Bold fonts indicate an improved or comparable performance compared to \protect\WM.}
    \centering
    \resizebox{\columnwidth}{!}{
    \begin{tabular}{lccccc}
    \hline
    Question & Count & Accuracy & Precision & Recall & F1 \\
    \hline
    Smooth or featured&8609&\textbf{0.948}&\textbf{0.944}&\textbf{0.948}&\textbf{0.945}\\
    Disc edge on &2986&\textbf{0.988}&\textbf{0.988}&\textbf{0.988}&\textbf{0.988}\\
    Has spiral arms &2788&0.910&0.915&0.910&0.912\\
    Bar &2178&\textbf{0.821}&\textbf{0.873}&\textbf{0.821}&\textbf{0.830}\\
    Bulge size &2182&0.774&0.923&0.774&0.833\\
    How rounded &5504&\textbf{0.936}&\textbf{0.936}&\textbf{0.936}&\textbf{0.936}\\
    Edge on bulge &427&\textbf{0.932}&\textbf{0.950}&\textbf{0.932}&\textbf{0.940}\\
    Spiral winding &1562&0.791&0.801&0.791&0.772\\
    Spiral arm count &1558&0.749&0.924&0.749&0.812\\
    Merging &7925&\textbf{0.873}&\textbf{0.881}&\textbf{0.873}&\textbf{0.835}\\
    \hline
    \end{tabular}}
    \label{tab: performance}
\end{table}

\section{Results}
\label{sec: results}
After training in the source domain and the target domain, we can obtain the expected probability $\rho_q^{m_q}$ of 34 morphology features for a given galaxy, which is calculated by equation (\ref{eq: alpha2q}) from the predicted Dirichlet distribution parameter $\alpha_{q}^{m_q}$. In this section, we first show the source domain model's performance and then the target domain model's.

\begin{table}
    \caption{Classical performance metrics for all the \num{3618} BMz images from the overlapping footprint between DECaLS and BMz. After filtering by volunteers' total votes, there remain 835 galaxies.} Each row represents a question.
    \centering
    \resizebox{\columnwidth}{!}{
    \begin{tabular}{lccccc}
    \hline
    Question & Count & Accuracy & Precision & Recall & F1 \\
    \hline
    \multicolumn{6}{c}{\textit{(a) Source domain model predict on BMz}} \\
    Smooth or featured&835&0.834&0.855&0.834&0.828\\
    Disc edge on &204&0.867&0.990&0.867&0.924\\
    Has spiral arms &165&0.878&0.980&0.878&0.925\\
    Bar &133&0.759&0.854&0.759&0.798\\
    Bulge size &129&0.744&0.840&0.744&0.781\\
    How rounded &179&0.905&0.908&0.905&0.903\\
    Edge on bulge &16&0.812&0.932&0.812&0.865\\
    Spiral winding &98&0.755&0.843&0.755&0.790\\
    Spiral arm count &79&0.835&0.849&0.835&0.842\\
    Merging &341&0.973&0.971&0.973&0.972\\
    \multicolumn{6}{c}{\textit{(b) Target domain model predict on BMz}} \\
    Smooth or featured&835&\textbf{0.875}&\textbf{0.879}&\textbf{0.875}&\textbf{0.872}\\
    Disc edge on &204&\textbf{0.926}&\textbf{0.990}&\textbf{0.926}&\textbf{0.957}\\
    Has spiral arms &165&\textbf{0.933}&\textbf{0.972}&\textbf{0.933}&\textbf{0.952}\\
    Bar &133&\textbf{0.834}&\textbf{0.864}&\textbf{0.834}&\textbf{0.849}\\
    Bulge size &129&\textbf{0.798}&\textbf{0.827}&\textbf{0.798}&\textbf{0.805}\\
    How rounded &179&0.877&0.902&0.877&0.878\\
    Edge on bulge &16&\textbf{0.875}&\textbf{0.937}&\textbf{0.875}&\textbf{0.900}\\
    Spiral winding &98&\textbf{0.795}&\textbf{0.842}&\textbf{0.795}&\textbf{0.812}\\
    Spiral arm count &79&\textbf{0.860}&\textbf{0.843}&\textbf{0.860}&\textbf{0.851}\\
    Merging &341&\textbf{0.967}&\textbf{0.969}&\textbf{0.967}&\textbf{0.968}\\
    \hline
    \end{tabular}}
    \label{tab: compare_uda}
\end{table}

\subsection{Classical performance metrics of source domain model: DECaLS galaxies}
\label{subsec: performance}
We first evaluate the performance of the source domain model, which also sets the upper bound classification performance of the target domain model. We compute accuracy, precision, recall, and F1 score in the validation set using \textsc{scikit-learn},\footnote{\url{https://pypi.org/project/scikit-learn}} as shown in Table~\ref{tab: performance}. Ground truth labels are determined by receiving more than 50\% of the volunteers' votes and at least a total of 30 votes\footnote{There is no essential difference in the results of our use of 30 and the use of 34 in \WM\ as a threshold for ground truth.} for the given questions. If a question has a dependency question, its answer to the dependency question must also meet the criterion. For example, the feature `No Bar' requires that both `Featured or Disc' and `Edge-on No' first be met with a vote fraction greater than 50\%, and that has more than 30 volunteers' votes. For model predictions on these galaxies with ground-truth morphology labels, we simply select the morphology features with $\rho_q^{m_q} > 0.5$. For a given morphology question $q$, if there is no feature with $\rho_q^{m_q}>0.5$ in the model prediction, we consider this prediction incorrect. We take almost the same performance calculation approach as in \WM. Most of these metrics perform similarly or better to \WM~(the metrics with improved or equivalent performance are in bold text), with poorer performance on the question `Has Spiral Arms', `Bulge Size', and `Spiral Arm Count'.

\subsection{Classical performance metrics of target domain model: BMz galaxies}
To evaluate the performance of our target domain model, we test \num{3618} BMz galaxies in the overlapping footprint so that they have ground truth morphology labels. These galaxies are explicitly excluded from the two-step training and validation sets, ensuring that they remained unknown to the models. For each question, we take the same performance metrics calculations as in the previous subsection and list them in Table~\ref{tab: compare_uda}.

To have a better visual evaluation of the target domain model's performance, we first directly apply the source domain model to these \num{3618} BMz galaxies and show the model performance in the upper part of Table~\ref{tab: compare_uda}. Compared with the source domain model on DECaLS galaxies, we find that the source domain model performance on BMz galaxies is significantly decreased on almost all questions. For example, the accuracy of `Smooth or Featured' decreases by approximately 11\%, `Bar' by approximately 7\%, `Spiral Winding' by around 4\%, and `Spiral Arm Count' by roughly 9\%. The degradation of the source domain model on BMz galaxies confirms that there is a data shift between DECaLS and BMz galaxies and the necessity of DA.

The performance of the model after our UDA method is shown in the lower part of Table~\ref{tab: compare_uda}. As can be seen, the global performance of the target domain model is significantly improved and becomes very close to the source domain model for most of the morphology features. For example, for the morphology label `Bar', the source domain model has 82.1\% on DECaLS and 75.9\% on BMz galaxies, respectively, while after DA, it achieves 83.4\%. Despite the success of our target domain model on most morphology labels, we also find it difficult to increase the performance of `How Rounded' and `Merging'.

For the morphology feature `Merging', it is very likely that the decrease in the UDA model performance is caused by fluctuations of the small number of test galaxies, since the performance of the source domain model on these BMz test galaxies is unexpectedly good (accuracy=97.3\%), even much higher than the source domain model on DECaLS galaxies (accuracy=87.3\%, Table~\ref{tab: performance}). The impact of the morphology feature `How Rounded' on classification performance remains uncertain; however, this reduction is considered an acceptable trade-off.

\begin{table}
    \caption{Classical performance metrics of our target domain model on all the BMz galaxies when treating prediction from \citet{walmsley_galaxy_2023} as true labels. Each row represents a question. We bold comparable ($\pm 0.05$) or better performance compared to source domain model on GZD galaxies.}
    \centering
    \resizebox{\columnwidth}{!}{
    \begin{tabular}{lccccc}
    \hline  
    Question & Count & Accuracy & Precision & Recall & F1 \\
    \hline
    Smooth or featured & 227262 & \textbf{0.966} & \textbf{0.966} & \textbf{0.966} & \textbf{0.966} \\
    Disc edge on & 54736 & 0.910 & 0.988 & 0.910 & 0.948 \\
    Has spiral arms & 38187 & \textbf{0.908} & \textbf{0.946} & \textbf{0.908} & \textbf{0.926} \\
    Bar & 25846 & \textbf{0.872} & \textbf{0.942} & \textbf{0.872} & \textbf{0.901} \\
    Bulge size & 30069 & \textbf{0.883} & \textbf{0.926} & \textbf{0.883} & \textbf{0.903} \\
    How rounded & 170625 & 0.887 & 0.912 & 0.887 & 0.894 \\
    Edge on bulge & 7518 & \textbf{0.930} & \textbf{0.962} & \textbf{0.930} & \textbf{0.945} \\
    Spiral winding & 15918 & \textbf{0.816} & \textbf{0.882} & \textbf{0.816} & \textbf{0.834} \\
    Spiral arm count & 14861 & \textbf{0.940} & \textbf{0.971} & \textbf{0.940} & \textbf{0.954} \\
    Merging & 219004 & 0.979 & 0.989 & 0.979 & 0.983 \\
    \hline
    \end{tabular}}
    \label{tab: compw23}
\end{table}

\subsubsection{Comparison with the results of \citet{walmsley_galaxy_2023}}
\label{sec: cp23}
Recently \cite{walmsley_galaxy_2023} released the newly collected volunteers' votes for \num{54716}\footnote{Core sample with votes of artifact < 5.} DESI-LIS galaxies ($m_\text{r} < 19$) (GZD-8, including BMz galaxies). They used all the GZ labels to fine-tune the  \texttt{Zoobot} model and give a prediction for \num{8700000} galaxies in DESI-LIS.
In this subsection, we use their predictions of all $m_r<17.77$ BMz galaxies (\num{227262}) for 34 morphology features as labels\footnote{\url{https://zenodo.org/records/8360385}} to compare consistency. For comparison with GZD-8 labels, we discuss in the Appendix~\ref{app: comp_gzd8}. We use the same performance metrics as in Sect.~\ref{subsec: performance}. We select the morphology feature with the highest probability as the predicted morphology feature and handle dependencies as before.

As shown in Table~\ref{tab: compw23}, we present the classical performance metrics of our target domain model's predictions, demonstrating strong consistency with the predictions from \cite{walmsley_galaxy_2023}. Specifically, seven morphology questions in the target domain exhibit comparable or better performance metrics compared to source domain model in the source domain (GZD), achieving an accuracy of 97.9\% for the best morphology question and 81.5\% for the worst. This consistency indicates that the target domain model does not exhibit significant bias under data shift. Additionally, our results confirm that the labels annotated by GZD-8 volunteers do not show obvious bias between GZD and BMz.

Specifically, seven morphology questions in the target domain exhibit comparable or better performance metrics compared to the source domain model in the source domain (GZD), achieving an accuracy of 97.9\% for the best morphology question and 81.5\% for the worst. This consistency indicates that the target domain model does not exhibit significant bias under data shift.

\begin{figure*}
\includegraphics[width=\textwidth]{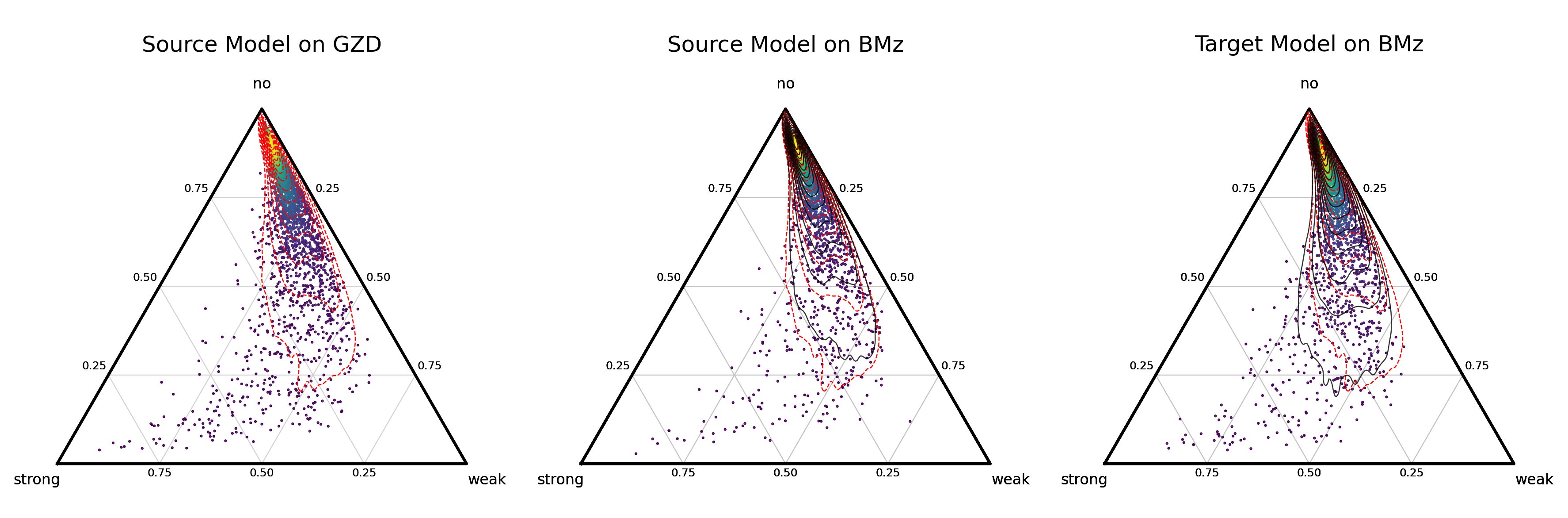}
    \caption{Expected probability $\hat{\rho}_q^{m_q}$ of the Dirichlet distribution of the model output visualised by probability simplex for the question `Bar', where the three vertices represent the corresponding three features, namely `Weak Bar', `No Bar', and `Strong Bar'. The scatter point is 1\% sampling from BMz galaxies. Each data point within a triangle represents the expected probability combinations of the features. To read the probability of a feature, draw a line parallel to its opposing side, and the intersection at the right side indicates the probability (bottom edge: `Strong Bar',  top right edge: `Weak Bar', top left edge: `No Bar'). The left, middle, and right panels show the cases for `the source model on DECaLS galaxies', `the source model on BMz galaxies' and `the target model on BMz galaxies', respectively. In each panel, the black contours show the number density distributions of the data points. In the middle and right panel, the red dashed contours are copies of the result of the source domain (left panel).}
\label{fig: dirichlet}.
\end{figure*}

\subsection{From DECaLS to BMz: morphology feature probability distribution}\label{subsec:from_to}
  
To further show the ability of our UDA model and to avoid fluctuations of performance metrics caused by limited test galaxies, we compare the predicted Dirichlet distribution of DECaLS and BMz galaxies from the source and target domain model for each morphology question, respectively. As we have introduced, the physical domains of DECaLS and BMz are the same as MGS, a perfect UDA model should predict the same probability distribution of the morphology features on all BMz galaxies as the source domain model on all DECaLS galaxies.

Specifically, we use probability simplex to visualise the Dirichlet distributions of our models' prediction (equation~(\ref{eq: alpha2q})).
We take the question `Bar' as an example, which has achieved the most significant enhancement after UDA (with a accuracy from 75.9\% to 83.4\%).
We plot the results of the source domain model, the application of the source domain model to the BMz galaxies, and the target domain model in the left, middle, and right panels of Fig.~\ref{fig: dirichlet}, respectively. As can be seen from the middle panel, when we directly apply the source domain model to the target domain, the proportion of the `Strong Bar' and `Weak Bar' galaxies are significantly underestimated, as evidenced by the incomplete coverage of the model prediction (black contours) to the source domain model results (red contours). After UDA, the distribution of bar morphology features of galaxies in the target domain is much closer to that of the source domain result (right panel). On the other hand, our target domain model anticipates a higher number of galaxies located at the central region of the probability simplex. The overabundance of galaxies in this region implies a higher fraction of BMz galaxies, so that our model cannot differentiate their bar features effectively, which is consistent with the fact that the image resolutions of the BMz galaxies are lower than those of the source domain (see Table~\ref{tab: survey}).

\begin{figure*}
\centering
\includegraphics[width=\textwidth]{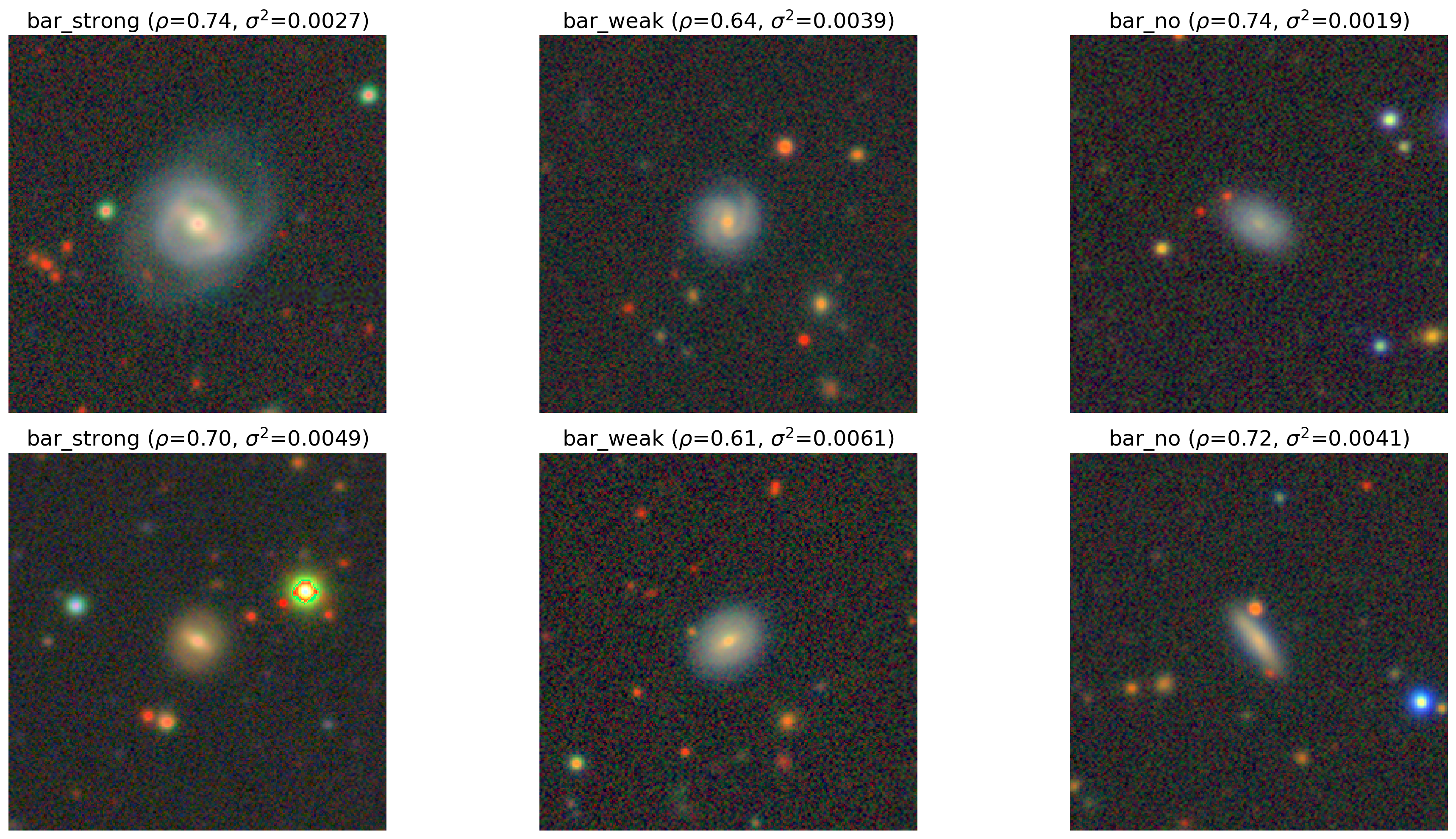}
        \caption{Examples of BMz galaxies are shown with `Strong Bar' (left), `Weak Bar' (middle), and `No Bar' features (right), respectively. All galaxies are selected with $\hat{\rho}_\text{bar}^{m_\text{bar}} > 0.5$. The galaxies in the top row have lower variance (the top 15\% in $\sigma^2$), while those in the bottom row have higher variance (the bottom 15\% in $\sigma^2$).}
    \label{fig: special_var_1}
\end{figure*}

\begin{figure*}
\centering
\includegraphics[width=\textwidth]{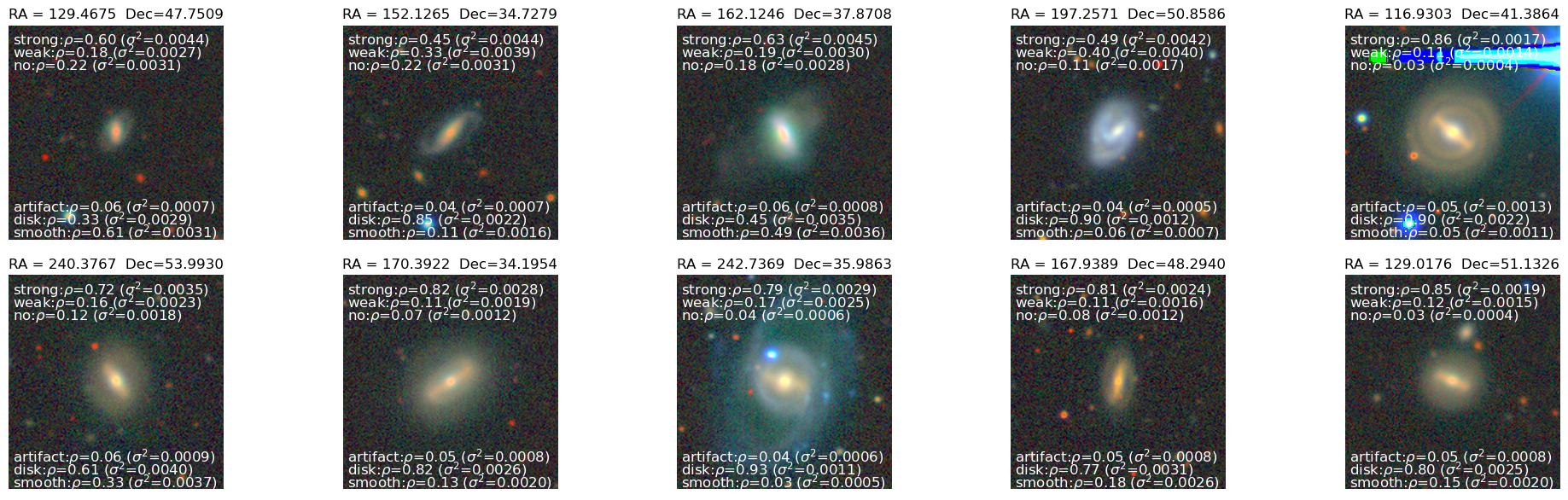}
        \caption{Example of BMz galaxies selected with `Strong Bar' features. The top row of galaxies are following a decision tree: $\hat{\rho}_\text{smooth or featured}^\text{featured or disc}$, $\hat{\rho}_\text{edge on}^\text{not edge on}$, $\hat{\rho}_\text{bar}^\text{strong bar}$ are larger than other features (the top 30\% $\sigma^2$), and while the galaxies at bottom are selected, simply selected with $\hat{\rho}_\text{bar}^\text{strong bar}$ is larger than other features (the top 30\% $\sigma^2$).}
    \label{fig: var_and_depend}
\end{figure*}
\section{Galaxy Morphology Catalogue}
\label{sec: use}
Our catalogue provides comprehensive classifications for 10 morphology questions across all \num{248088} BMz galaxies. It lists the predicted Dirichlet distribution parameter $\bar{\alpha}_{q}^{m_q}$, averaged over 100 instances of MC Dropout \citep{gal_dropout_2016}, for each morphology feature $m_q$, and the associated question $q$. The columns for $\bar{\alpha}_{q}^{m_q}$ are named using the format \texttt{\{question\}\_\{answer\}\_alpha}. 
For example, the column \texttt{bar\_no\_alpha} represents $\bar{\alpha}_\text{bar}^\text{no bar}$. Additionally, the catalogue includes the expected probability $\hat{\rho}_q^{m_q}$, which indicates the likelihood that each galaxy corresponds to each morphology feature. This probability is related to $\bar{\alpha}_{q}^{m_q}$ through the following equation:

\begin{equation}
\label{eq: alpha2q}
\hat{\rho}_q^{m_q}= \dfrac{\bar{\alpha}_q^{m_q}}{\hat{\alpha}_0}, \rm{with} \,\, \hat{\alpha}_0=\sum_{i=1}^{K} \bar{\alpha}_q^{m_q}\,.
\end{equation}
The column name format for $\hat{\rho}_q^{m_q}$ is \texttt{\{question\}\_\{answer\}\_prob}. Additionally, the catalogue lists the variance of the probability $\hat{\rho}_q^{m_q}$, whose column name format is \texttt{\{question\}\_\{answer\}\_var} and is calculated by
\begin{equation}
\label{eq: var}
\sigma^2=\dfrac{\hat{\rho}_q^{m_q}\left(1-\hat{\rho}_q^{m_q}\right)}{\hat{\alpha}_{0}+1}, 
\end{equation}

A straightforward approach to filtering the morphology features of interest $(q,m_q)$ is to apply a predefined probability threshold $\hat{\rho}_q^{m_q}$ or choose the maximum $\hat{\rho}_q^{m_q}$ for a given question $q$. For example, we can filter galaxies with `Strong Bar', `Weak Bar', and `No Bar' with the corresponding $\rho_\text{bar}^{m_\text{bar}}$ as shown in Fig.~\ref{fig: special_var_1}.

On the other hand, the $\hat{\rho}_q^{m_q}$ obtained from a Dirichlet distribution with low $\bar{\alpha}_{q}^{m_q}$ values can be the same as that with high $\bar{\alpha}_{q}^{m_q}$ values, despite higher uncertainties (equation~(\ref{eq: var})). As an illustration, we randomly select galaxies with `Strong Bar', `Weak Bar', and `No Bar' with similar $\hat{\rho}_\text{bar}^{m_\text{bar}}$  values but across different uncertainty (variance) ranges, respectively. The results are shown in Fig.~\ref{fig: special_var_1}.  It is evident that the galaxies in the bottom row with a larger variance in predicted probability $\hat{\rho}_q^{m_q}$ (characterised by low values $\bar{\alpha}_{q}^{m_q}$) show more ambiguities in the bar feature than those of the corresponding galaxies in the top row (lower $\sigma^2$). Therefore, to confidently select a galaxy with a particular morphology feature, we recommend considering both $\hat{\rho}_q^{m_q}$ and its $\sigma^2$.

So far, we have simplified the selection of the morphology features of the galaxy $(q, m_q)$ into a single step using $\hat{\rho}_q^{m_q}$ and $\sigma^2$, following our concomitant training methodology for all morphology questions (Sect.~\ref{subsec: network}). In contrast, for the training galaxies within the GZD-5 project, each volunteer's vote follows a decision tree structure. For example, volunteers only vote on the `Bar' feature for galaxies classified as `Featured or Disc' and not for `Edge-on Yes' galaxies. This means that for inquiries about sub-features within the decision tree, such as `Has Spiral Arms', `Strong Bar', or `Tight Spiral Arms', the training dataset contains more votes for galaxies that have successfully passed the preceding higher-level question, such as being classified as `Featured or Disc' galaxies.

Therefore, to conservatively select a galaxy morphology feature like GZD-5 volunteers, we can follow the same decision tree used for DECaLS and BMz galaxies. An example of this approach is shown in Fig.~\ref{fig: var_and_depend}, where a sample of strong bar galaxies selected using simple $\hat{\rho}_\text{bar}^\text{strong}$ and $\sigma^2$ criteria is compared with those selected using a complete decision tree. As seen, the galaxies with bars selected using the complete decision tree (bottom panel) are confidently disc galaxies with strong bar features. For galaxies selected solely on the basis of bar feature criterion (top row), most are disc galaxies with strong bar features. However, it is interesting to note that this group also includes some elliptical-like galaxies with bar features (e.g. the 1st column of the top row). The features of the bars in the elliptical-shaped light profiles may indicate S0-type galaxies \citep{hubble_extragalactic_1926}, which has been explored in previous studies (e.g. \citealt{dullo_complex_2016,tahmasebzadeh_schwarzschild_2023,tsvetkov_star_2024}).

In summary, the choice between using a single condition or a decision tree to select galaxy morphology features depends on the specific research objectives and user requirements.

\section{Conclusion}
\label{sec: conclusion}
In this study, we explore a UDA method that can fine-tune the detailed galaxy morphology classification model from one survey to another without collecting new labels or common galaxies when they are in the same physical domain. 

We first trained a model as a foundation on DESI-LIS DECaLS images and the votes of the GZD-5 volunteers (as a source domain model), which can predict the Dirichlet distribution of the detailed morphology features of galaxies of $z<0.15$, $m_r<17.77$ and Petrosian radius$>3$ arcsec and has a performance comparable to the previous study of \WM. We tend to apply this neural network to galaxies of the same physical domain in the BMz survey to increase the sample of galaxies with detailed classifications of morphology features. We find that the data shift between DECaLS and BMz datasets (e.g. resolution, noise) results in a performance decrease when the source domain model is directly applied on BMz galaxies. To accomplish the data shift from DECaLS to BMz, we fine-tuned the source domain model on BMz galaxies by the UDA method. The fine-tuned target domain model achieves an improvement in most questions and mitigated the bias between the source and target domain. We release a catalogue of \num{248088} detailed galaxies morphology classification in the DESI-LIS BMz survey and the corresponding model's weight of both domains. This catalogue have high consistency with the prediction from \cite{walmsley_galaxy_2023}, which is fine-tuned on additional visual inspection from GZD-8. For the sake of completeness, the source domain model predictions on \num{345650} DECaLS galaxies are also released for comparison. For the galaxy morphology label, each galaxy contains the expected probabilities and variances for 34 morphology feature answers of 10 different morphology questions. To select a sample of galaxies with a specific morphology feature, one may use a single morphology label or a combination of multiple labels, depending on the question being discussed.

This study complements the \texttt{Zoobot} series study and addresses the problem of data shift with a label-free strategy. Our study provides an efficient way of migrating galaxy morphology classification labels from one survey to another, which can be easily adapted for future astronomical surveys, such as CSST \citep{gong_cosmology_2019}, Euclid \citep{euclid_collaboration_euclid_2022}, and LSST \citep{ivezic_lsst_2019}. However, it should be emphasized that this UDA algorithm relies on the assumption that the physical properties inherent in two different domains should have the same distribution, as the two samples of galaxies in this study, which are both low-redshift bright galaxies. To extend the galaxy morphology classification to different physical domains, e.g. fainter galaxies, our UDA method needs to be further explored. For example, during the alignment of the morphology feature embeddings extracted from the neural network (equation~(\ref{eq:uda_loss})), it is necessary to distinguish which features in the target domain are of the same origin as the source domain and which are new features. 

\section*{Acknowledgements}
This work was supported by research grants from the National Key R\&D Program of China (No. 2022YFF0503402, 2019YFA0405501), National Natural Science Foundation of China (No. 12073059 \& 12141302 ), Shanghai Academic/Technology Research Leader (22XD1404200) and the China Manned Space Project with NO. CMS-CSST-2021-A07. We acknowledge the PixInsight software for the analysis of DESI-LIS images.

The Legacy Surveys consist of three individual and complementary projects: the Dark Energy Camera Legacy Survey (DECaLS; Proposal ID \#2014B-0404; PIs: David Schlegel and Arjun Dey), the Beijing-Arizona Sky Survey (BASS; NOAO Prop. ID \#2015A-0801; PIs: Zhou Xu and Xiaohui Fan), and the Mayall $z$-band Legacy Survey (MzLS; Prop. ID \#2016A-0453; PI: Arjun Dey). DECaLS, BASS and MzLS together include data obtained, respectively, at the Blanco telescope, Cerro Tololo Inter-American Observatory, NSF’s NOIRLab; the Bok telescope, Steward Observatory, University of Arizona; and the Mayall telescope, Kitt Peak National Observatory, NOIRLab. Pipeline processing and analyses of the data were supported by NOIRLab and the Lawrence Berkeley National Laboratory (LBNL). The Legacy Surveys project is honored to be permitted to conduct astronomical research on Iolkam Du’ag (Kitt Peak), a mountain with particular significance to the Tohono O’odham Nation.

NOIRLab is operated by the Association of Universities for Research in Astronomy (AURA) under a cooperative agreement with the National Science Foundation. LBNL is managed by the Regents of the University of California under contract to the U.S. Department of Energy. 

This project used data obtained with the Dark Energy Camera (DECam), which was constructed by the Dark Energy Survey (DES) collaboration. Funding for the DES Projects has been provided by the U.S. Department of Energy, the U.S. National Science Foundation, the Ministry of Science and Education of Spain, the Science and Technology Facilities Council of the United Kingdom, the Higher Education Funding Council for England, the National Center for Supercomputing Applications at the University of Illinois at UrbanaChampaign, the Kavli Institute of Cosmological Physics at the University of Chicago, Center for Cosmology and Astro-Particle Physics at the Ohio State University, the Mitchell Institute for Fundamental Physics and Astronomy at Texas A\&M University, Financiadora de Estudos e Projetos, Fundacao Carlos Chagas Filho de Amparo, Financiadora de Estudos e Projetos, Fundacao Carlos Chagas Filho de Amparo a Pesquisa do Estado do Rio de Janeiro, Conselho Nacional de Desenvolvimento Cientifico e Tecnologico and the Ministerio da Ciencia, Tecnologia e Inovacao, the Deutsche Forschungsgemeinschaft and the Collaborating Institutions in the Dark Energy Survey. The Collaborating Institutions are Argonne National Laboratory, the University of California at Santa Cruz, the University of Cambridge, Centro de Investigaciones Energeticas, Medioambientales y Tecnologicas-Madrid, the University of Chicago, University College London, the DES-Brazil Consortium, the University of Edinburgh, the Eidgenossische Technische Hochschule (ETH) Zurich, Fermi National Accelerator Laboratory, the University of Illinois at Urbana-Champaign, the Institut de Ciencies de l’Espai (IEEC/CSIC), the Institut de Fisica d’Altes Energies, Lawrence Berkeley National Laboratory, the Ludwig Maximilians Universitat Munchen and the associated Excellence Cluster Universe, the University of Michigan, NSF’s NOIRLab, the University of Nottingham, the Ohio State University, the University of Pennsylvania, the University of Portsmouth, SLAC National Accelerator Laboratory, Stanford University, the University of Sussex, and Texas A\&M University. 

BASS is a key project of the Telescope Access Program (TAP), which has been funded by the National Astronomical Observatories of China, the Chinese Academy of Sciences (the Strategic Priority Research Program "The Emergence of Cosmological Structures" Grant \# XDB09000000), and the Special Fund for Astronomy from the Ministry of Finance. The BASS is also supported by the External Cooperation Program of Chinese Academy of Sciences (Grant \# 114A11KYSB20160057), and Chinese National Natural Science Foundation (Grant \# 12120101003, \# 11433005). 

The Legacy Survey team makes use of data products from the Near-Earth Object Wide-field Infrared Survey Explorer (NEOWISE), which is a project of the Jet Propulsion Laboratory/California Institute of Technology. NEOWISE is funded by the National Aeronautics and Space Administration. 

The Legacy Surveys imaging of the DESI footprint is supported by the Director, Office of Science, Office of High Energy Physics of the U.S. Department of Energy under Contract No. DE-AC0205CH1123, by the National Energy Research Scientific Computing Center, a DOE Office of Science User Facility under the same contract; and by the U.S. National Science Foundation, Division of Astronomical Sciences under Contract No. AST-0950945 to NOAO.

\section*{Data Availability}
\label{sec: public}
This work uses morphology labels from \cite{walmsley_galaxy_2023}, available at \url{https://doi.org/10.5281/zenodo.4573248}.
Our detailed morphology classification of \num{248088} galaxies on the DESI-LIS (BMz) is available at \url{https://doi.org/10.5281/zenodo.10579386}.
Our code will be available on \url{https://github.com/Rh-YE/ai4galmorph_desi} for understanding the algorithm. We sincerely hope that you can make full use of the data for further analysis.


\bibliographystyle{mnras}
\bibliography{Ye2023} 
\bsp	

\appendix
\section{Comparison with GZD-8 labels}
\label{app: comp_gzd8}
In Sect.\ref{sec: cp23}, we discussed the consistency between our prediction and the findings of \cite{walmsley_galaxy_2023}. Since the GZD-8 labels include galaxies in the BMz region, we have the opportunity to discuss possible biases of human visual inspection. Cross-matching our predictions with the GZD-8 labels results in only \num{552} galaxies, since most newly collected galaxies in the BMz region are faint, with magnitudes in the range $17.77 < m_r < 19$. We calculate the classical performance metrics as the same in Sect.~\ref{subsec: performance}. As shown in Table~\ref{tab: comp_votes}, we find a comparable performance with the bottom of Table~\ref{tab: compare_uda}. However, detailed features like `Spiral winding' and `Spiral arm count' show obvious inconsistencies of more than 12\% accuracy difference. It shows preliminary evidence that volunteers' votes on the detailed structure may have bias due to the resolution difference between GZD and BMz.

\begin{table}
    \caption{Classical performance metrics of our target domain model on all the BMz galaxies with GZD-8 labels. Each row represents a question.}
    \centering
    \resizebox{\columnwidth}{!}{
    \begin{tabular}{lccccc}
    \hline
    Question & Count & Accuracy & Precision & Recall & F1 \\
    \hline
    Smooth or featured&552&0.900&0.899&0.900&0.899\\
    Disc edge on &135&0.962&0.963&0.962&0.963\\
    Has spiral arms &125&0.888&0.883&0.888&0.874\\
    Bar &125&0.768&0.772&0.768&0.769\\
    Bulge size &125&0.800&0.816&0.800&0.802\\
    How rounded &361&0.886&0.889&0.886&0.886\\
    Edge on bulge &22&1.000&1.000&1.000&1.000\\
    Spiral winding &99&0.656&0.692&0.656&0.666\\
    Spiral arm count &99&0.747&0.747&0.747&0.736\\
    Merging &552&0.929&0.941&0.929&0.926\\
    \hline
    \end{tabular}}
    \label{tab: comp_votes}
\end{table}

\section{Strong Bar sample}
We randomly select some `Strong Bar' galaxies from our target domain model prediction in the BMz galaxies.
\begin{figure}
\centering
\includegraphics[width=\columnwidth]{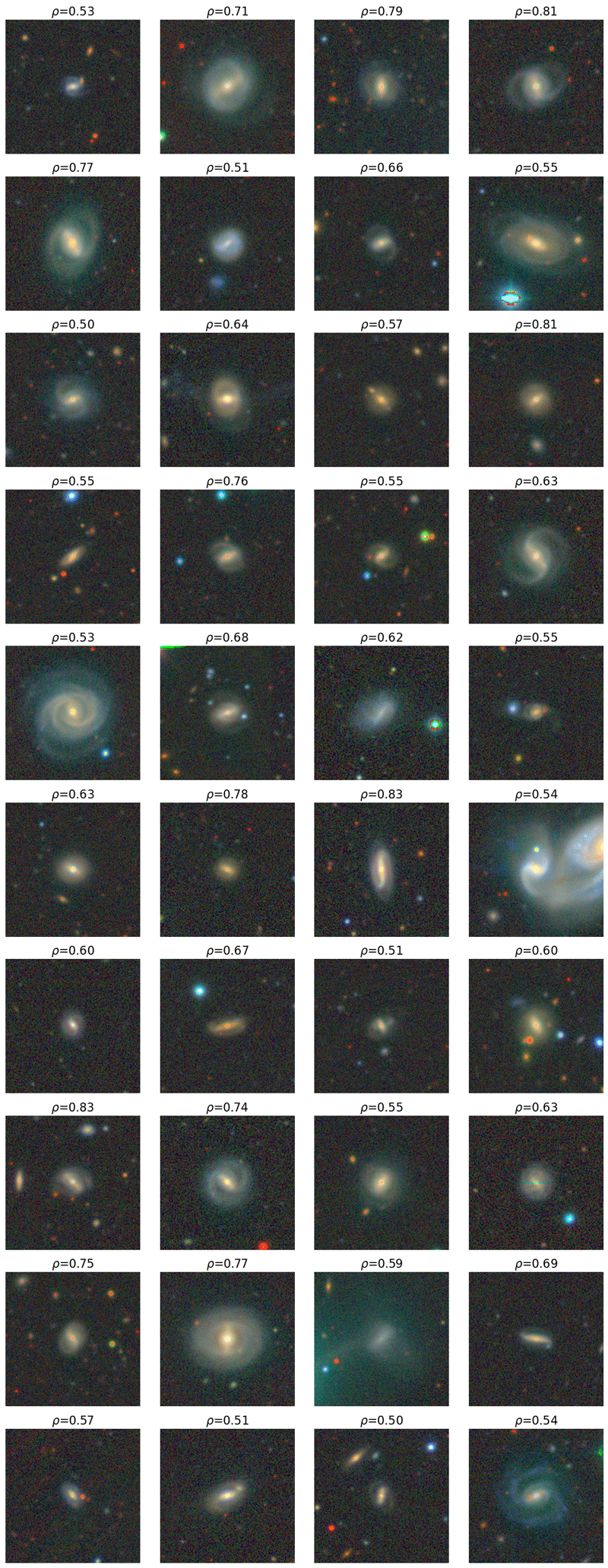}
        \caption{Example of BMz galaxies randomly selected by $\hat{\rho}_\text{bar}^{\text{strong bar}}>0.5$.}
    \label{fig: bar_sample}
\end{figure}

\label{lastpage}
\end{document}